\documentclass{PoS}

\usepackage{epsfig}

\title{Improvement of quark propagator estimation through domain 
decomposition}

\ShortTitle{Improvement of quark propagator estimation through domain 
decomposition}

\author{\speaker{T.~Burch} and C.~Hagen\\

  Institut f\"ur Theoretische Physik\\
  Universit\"at Regensburg\\
  D-93040 Regensburg, Germany.\\

  E-mail: \email{tommy.burch@physik.uni-r.de}\\
}

\abstract{
Applying domain decomposition to the lattice Dirac operator 
and the associated quark propagator, we arrive at expressions which, 
with the proper insertion of random sources therein, can provide 
improvement to the estimation of the propagator. 
Schemes are considered for both open and closed (or loop) propagators. 
In the end, our technique for improving open contributions is similar 
to the ``maximal variance reduction'' approach of Michael and Peisa, 
but contains the advantage, especially for improved actions, of 
dealing directly with the Dirac operator. 
Using these improved open propagators for the Chirally Improved 
operator, we present preliminary results for the static-light meson 
spectrum.
}

\FullConference{XXIVth International Symposium on Lattice Field Theory\\
                23-28 July 2006\\
                Tucson, Arizona, USA}

\begin{document}

\section{Introduction}\label{SectIntroduction}

We present a method \cite{Burch:2006mb} for improving the estimation of quark 
propagators between different domains of the lattice. 
Our method turns out to be similar to that of ``maximal variance reduction'' 
(MVR) \cite{Michael:1998sg}. 
However, it contains the advantage that one can work directly with the chosen 
lattice Dirac operator. 
In the following sections, we present our method and some first results for 
static-light mesons, where the Chirally Improved (CI) \cite{CI_action}, 
light quark propagator is calculated with the improved estimator.

\section{The method}\label{SectMethod}

Decomposing the lattice into two distinct regions, 
the full Dirac matrix can be written in terms of submatrices 
\begin{equation}
  M = \left(
    \begin{array}{cc}
      M_{11} & M_{12} \\
      M_{21} & M_{22}
    \end{array} \right) \, ,
\end{equation}
where $M_{11}$ and $M_{22}$ connect sites within a region and 
$M_{12}$ and $M_{21}$ connect sites from the different regions. 
We can also write the propagator in this form: 
\begin{equation}
  M^{-1} = P = \left(
    \begin{array}{cc}
      P_{11} & P_{12} \\
      P_{21} & P_{22}
    \end{array} \right) \, .
\end{equation}

We consider a set of random sources, $\chi^n$ ($n=1,...,N$), 
and the corresponding resultant vectors, $\eta^n=P\chi^n$, to derive useful 
expressions for our technique. 
Reconstructing the sources in one region, $\chi_1^n$, from the solution 
vectors everywhere, $\eta^n$, we may write 
\begin{equation}
  \chi_1^n = M \eta^n = M_{11} \eta_1^n + M_{12} \eta_2^n \, .
\end{equation}
If we now apply the inverse of the matrix within one region, we have 
\begin{equation}
  M_{11}^{-1} \chi_1^n = \eta_1^n + M_{11}^{-1} M_{12} \eta_2^n \, .
\end{equation}
This can be solved for $\eta_1^n$ and substituted back into the naive 
estimator of the propagator between the two regions 
(repeated source indices, $n$, are summed over): 
\begin{eqnarray}
  P_{12} &\approx& \frac1N \eta_1^n \chi_2^{n\dagger} \nonumber \\
  &\approx& \frac1N \left[ M_{11}^{-1} \left( \chi_1^n - M_{12} \eta_2^n \right) \right] \chi_2^{n\dagger} \nonumber \\
  &\approx& - \frac1N \left( M_{11}^{-1} M_{12} \eta_2^n \right) \chi_2^{n\dagger} \, ,
\end{eqnarray}
where in the last line we eliminate the first term due to the fact that 
we expect $\lim_{N\to\infty}\chi_1^n \chi_2^{n\dagger} = 0$. 
Writing out the full expression, we obtain 
\begin{eqnarray}
  \label{postdiluting}
  P_{12} &\approx& - \frac1N M_{11}^{-1} M_{12} P \chi_2^n \chi_2^{n\dagger} \nonumber \\
  &=& - M_{11}^{-1} M_{12} P_{22} \, ,
\end{eqnarray}
where the second line is an exact expression, showing that one 
can relate elements of different regions of $P=M^{-1}$ via the inverse of a 
submatrix of $M$. 
This is nothing new. After all, $P_{22}$ is the Schur complement of 
$M_{11}^{-1}$. 
But the lesson learned up to this point is that we need no sources in 
one of the two regions.

Looking again at Eq.\ (\ref{postdiluting}), one can see that we need not make 
the approximation 
$P_{22} \approx \frac1N P \chi_2^n \chi_2^{n\dagger}$. 
Instead, we can place the approximate Kronecker delta between the $M_{12}$ 
and $P_{22}$: 
\begin{eqnarray}
  \label{conn_estimator}
  P_{12} &\approx& - \frac1N M_{11}^{-1} M_{12} \chi_2^n \chi_2^{n\dagger} P \nonumber \\
  &\approx& - \frac1N \left( M_{11}^{-1} M_{12} \chi_2^n \right) \left( \gamma_5 P \gamma_5 \chi_2^{n} \right)^\dagger \nonumber \\
  &\approx& - \frac1N \psi_1^n \phi_2^{n\dagger} \, ,
\end{eqnarray}
where we have used the $\gamma_5$-hermiticity of the propagator. 
One can see from the form of the vector $\psi_1^n$ in the next to last line 
that we only need sources which ``reach'' region 1 via one application of 
$M$. 
Also, one can use all points in one region for the source and all points in 
the other region for the sink.

\begin{figure}
\begin{center}
\includegraphics*[width=6cm]{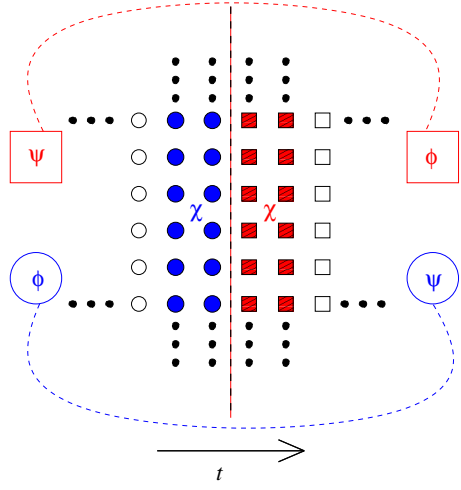}
\end{center}
\caption{
Depiction of one of the boundaries and the surrounding sources ($\chi$) 
which we use to construct the two estimates of the CI quark propagator 
between regions of equal volume. 
Colors and shapes indicate which source region contributes 
to the signal in the resultant vectors. 
The $\psi$'s are calculated using only one of the source regions, while 
the $\phi$'s use both.
}
\label{sources}
\end{figure}

For our first attempt of using this method, we use equal volumes for the two 
regions and place sources next to the boundaries in both regions 
(see Fig.\ \ref{sources}). Although this choice may not be ideal, we do 
perform inversions for each spin component separately (spin dilution) and, 
since our sources occupy all relevant time slices surrounding 
the boundaries, we actually obtain two independent estimates of the quark 
propagator between the two regions: 
\begin{equation}
  \label{conn_estimator2}
  - \frac1N \psi_1^n \phi_2^{n\dagger} \approx P_{12} \approx 
  - \frac1N \gamma_5 \phi_1^n \psi_2^{n\dagger} \gamma_5 \, .
\end{equation}

So our method is very similar to that of MVR, except 
for the fact that we can work directly with $M$, instead of $M^\dagger M$. 
This is better since it is less problematic to invert $M$ due to it having a 
better condition number than $M^\dagger M$ \cite{Michael:1998sg}. 
Also, the sources need only occupy enough time slices to connect them 
with the other region via $M$, rather than $M^\dagger M$. 
These are the same number of sources for Wilson-like operators, where 
$M^\dagger M$, like $M$, only extends one time slice. However, for many 
other improved operators (like CI) this can reduce the number of 
necessary source time slices by a factor of 2. 
On top of this, although the nature of the lattice Dirac operator may 
dictate what is the ideal domain decomposition, it does not otherwise 
restrict the choice of regions or the use of this method 
(e.g., it is even possible to use the Overlap operator).

For expressions and first results relevant to propagators which return to 
the same region (e.g., closed, or loop, propagators) we point the reader to 
our lengthier publication on the subject \cite{Burch:2006mb} and move on to 
our application for the open propagators.

\section{Static-light mesons}

For our meson source and sink operators, we use bilinears of 
the form: 
\begin{equation}
  \bar{Q} \, O(\Gamma,D_i,\vec D^2,S) \, q \, ,
\end{equation}
where $S$ is a gauge-covariant (Jacobi) smearing operator \cite{jacobi} and 
$\vec D$ is the covariant derivative. 
For our basis of light-quark spatial wavefunctions, we use three different 
amounts of smearing and apply 0, 1, and 2 covariant Laplacians to these: 
\begin{equation}
  q' = \; S_8 \, q \; , \; \vec D^2 S_{12} \, q \; , \; 
  \vec D^2 \vec D^2 S_{16} \, q \; ,
\end{equation}
where the subscript on the smearing operator denotes the number of smearing 
steps; all are applied with the same weighting factor of $\kappa_{sm}=0.2$. 
So we have a relatively narrow, approximately Gaussian distribution, along 
with wider versions which exhibit one and two radial nodes, due to the 
application of the Laplacians. 
We point out that, thus far, we have not altered the quantum numbers of the 
meson source since both the smearing and Laplacians treat all spatial 
directions the same (i.e., they are scalar operations).
In order to create mesons of different quantum numbers, we use these 
light-quark distributions together with the operators shown in 
Table \ref{sl_ops} (see, e.g., Ref.\ \cite{Michael:1998sg}).

\begin{table}
\begin{center}
\begin{tabular}{lcc} \hline
oper. & $J^P$ & $\bar{Q} \, O(\Gamma,D_i) \, q'$ \\ \hline
$S$ & $0^-,1^-$ & $\bar{Q} \, \gamma_5 \, q'$ \\
$P_-$ & $0^+,1^+$ & $\bar{Q} \, \sum_i \gamma_i D_i \, q'$ \\
$P_+$ & $1^+,2^+$ & $\bar{Q} \, (\gamma_1 D_1 - \gamma_2 D_2) \, q'$ \\
$D_\pm$ & $1^-,2^-,3^-$ & $\bar{Q} \, \gamma_5(D_1^2 - D_2^2) \, q'$ \\ \hline
\end{tabular}
\caption{
Static-light meson operators.
\label{sl_ops}
}
\end{center}
\end{table}

Inserting the estimated and static propagators in the meson correlators we 
have 
\begin{eqnarray}
  C_{ij}(t) &=& \left\langle 0 \left| (\bar Q \, O_j \, q)_t \; 
      (\bar q \, \bar O_i \, Q)_0 \right| 0 \right\rangle \nonumber \\
  &=& \left\langle \sum_x \mbox{Tr} \left[ 
      \frac{1+\gamma_4}{2}\prod_{i=0}^{t-1} U_4^\dagger(x+i\hat{4}) \, 
      O_j P_{x+t\hat{4},x} \bar O_i 
    \right] \right\rangle_{\{U\}} \, .
\end{eqnarray}
The static quark is propagated through products of links in the time 
direction and has a fixed spin $(1+\gamma_4)/2$. The estimated propagator 
$P_{x+t\hat{4},x}$ is of the form of Eq.\ (\ref{conn_estimator2}). 
Thus, all points within region 1 ($N_s^3N_t/2$ of them) can act as the 
source location $x$, just so long as $t$ is large enough to have the sink 
location $x+t\hat{4}$ in region 2. 
Note that we now have subscripts on the source and sink operators to denote 
which light-quark distribution is being used. 
We create all such combinations and thus have a $3\times3$ matrix of 
correlators for each of the operators in Table \ref{sl_ops}.

Following the work of Michael \cite{Mi85} and 
L\"uscher and Wolff \cite{LuWo90}, 
we use this cross-correlator matrix in a variational approach to separate 
the different mass eigenstates. 
We must therefore solve the generalized eigenvalue problem 
\begin{equation}
  \label{gep}
  C(t) \vec v^{(k)} = \lambda^{(k)}(t,t_0) \, C(t_0) \vec v^{(k)} \; ,
\end{equation}
in order to obtain the following eigenvalues: 
\begin{equation}
  \label{eigenvalues}
  \lambda^{(k)}(t,t_0) \propto e^{-t \, M_k} 
  [1+O(e^{-t \, \Delta M_k})] \, ,
\end{equation}
where $M_k$ is the mass of the $k$th state and $\Delta M_k$ is the 
mass-difference to the next state. 
For large enough values of $t$, each eigenvalue should then correspond to a 
unique mass state, requiring only a single-exponential fit.

Variational approaches have seen much use recently in lattice QCD, 
especially for extracting excited hadron masses and we point the reader 
to the relevant literature in \cite{exc_had}.

We create our cross-correlator matrices on two sets of gluonic 
configurations: 100 quenched and 74 dynamical, each with $12^3\times24$ 
lattices sites. 
The quenched configurations have a lattice spacing of $a \approx 0.15$ fm 
($a^{-1}\approx1330$ MeV) and a spatial extent of $L\approx1.8$ fm. 
The dynamical set \cite{Lang:2005jz} has 2 flavors of CI sea quarks (with 
$M_{\pi,\mbox{sea}}\approx500$ MeV), $a \approx 0.115$ fm 
($a^{-1}\approx1710$ MeV), and $L\approx1.4$ fm. 
We use 12 random spin-color vectors as sources for the light-quark 
propagator estimation. Spin-diluted, this gives us 48 separate sources for 
the inversions (one in the full volume, $\phi$, and two in the subregions, 
$\psi$; see Eqs.\ (\ref{conn_estimator}) and (\ref{conn_estimator2})). 
We perform inversions for 4 different quark masses: 
$am_q=0.02$, 0.04, 0.08, 0.10.

\begin{figure}
\begin{center}
\includegraphics*[width=3.7cm]{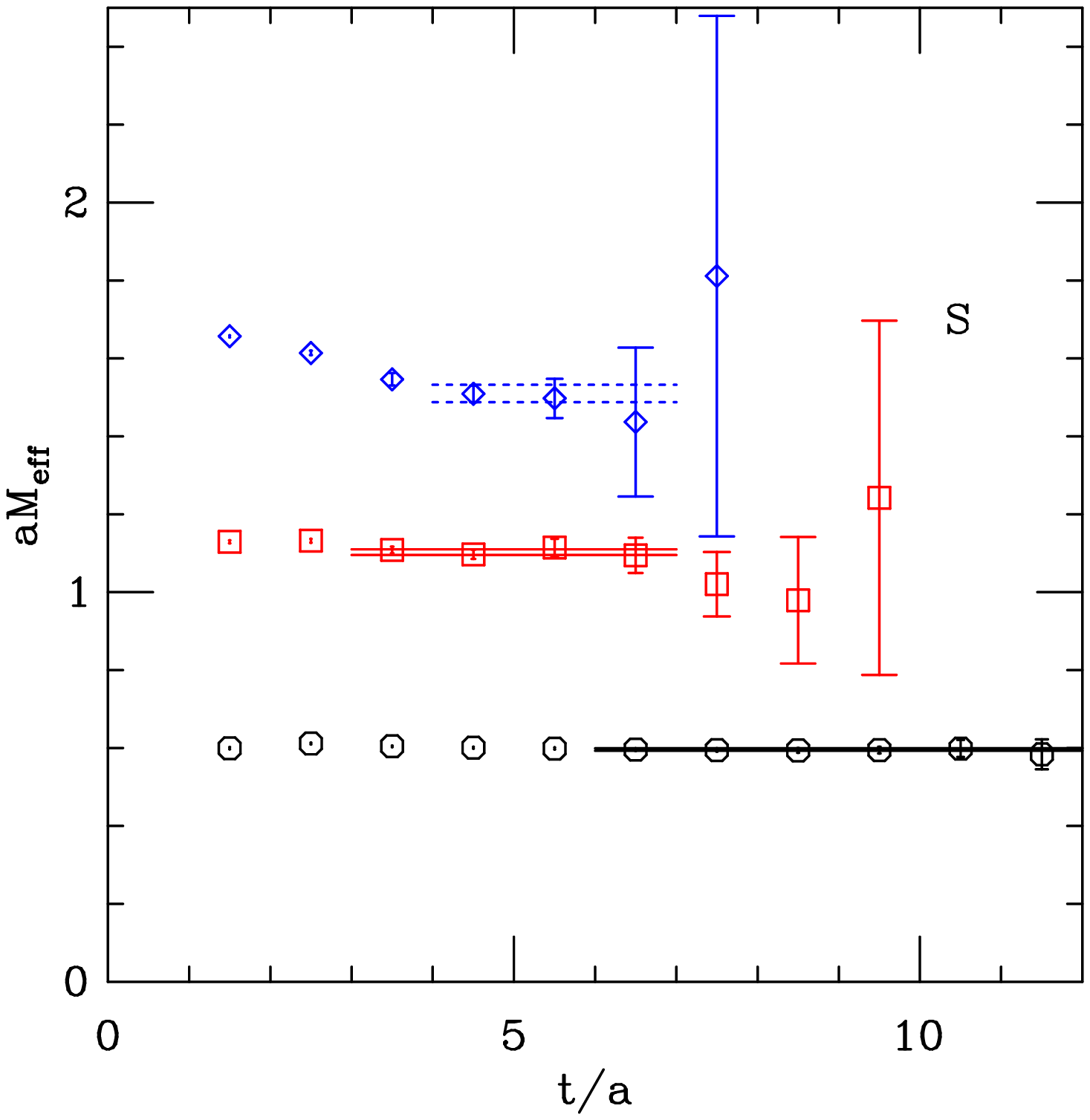}
\includegraphics*[width=3.7cm]{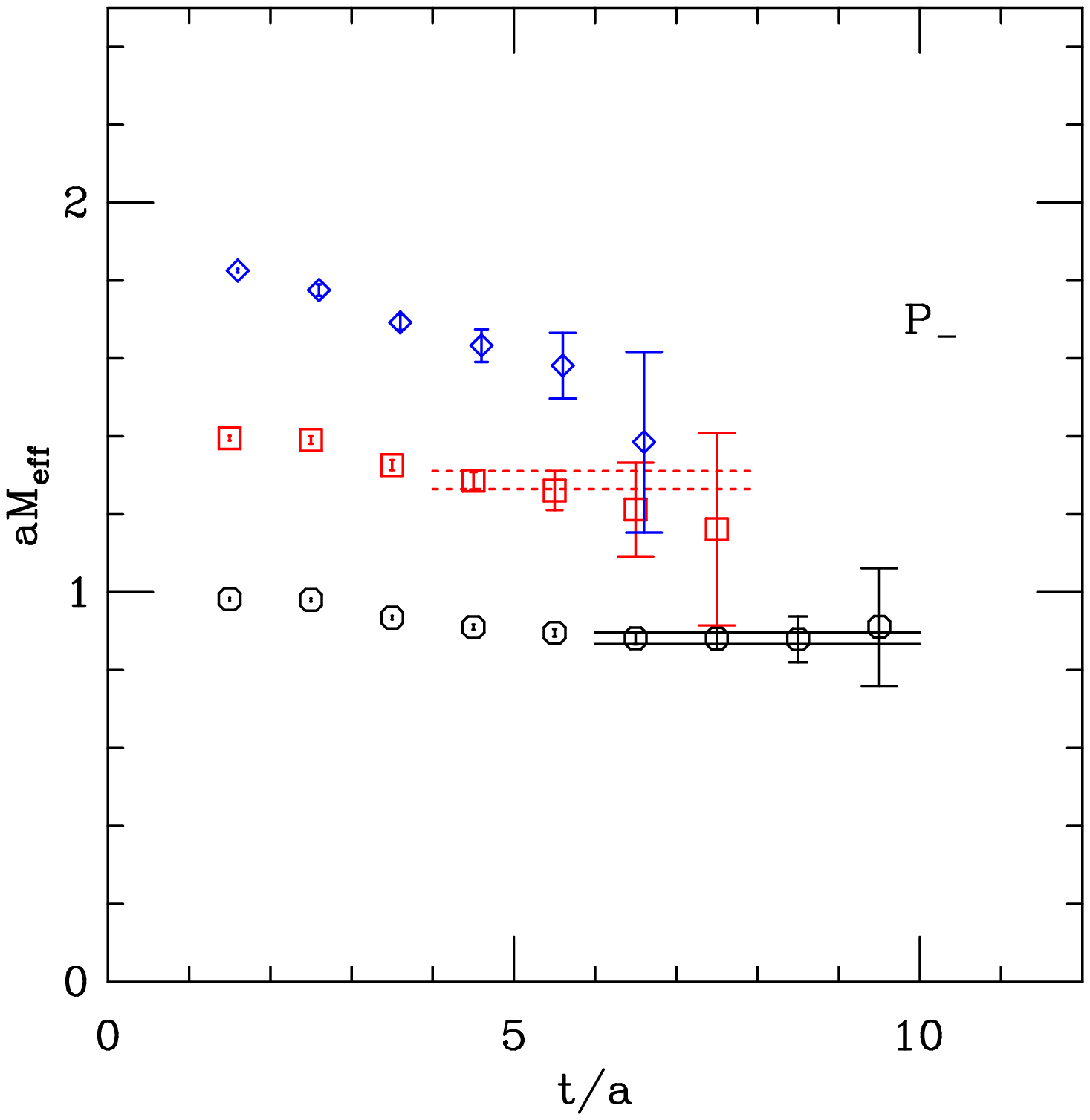}
\includegraphics*[width=3.7cm]{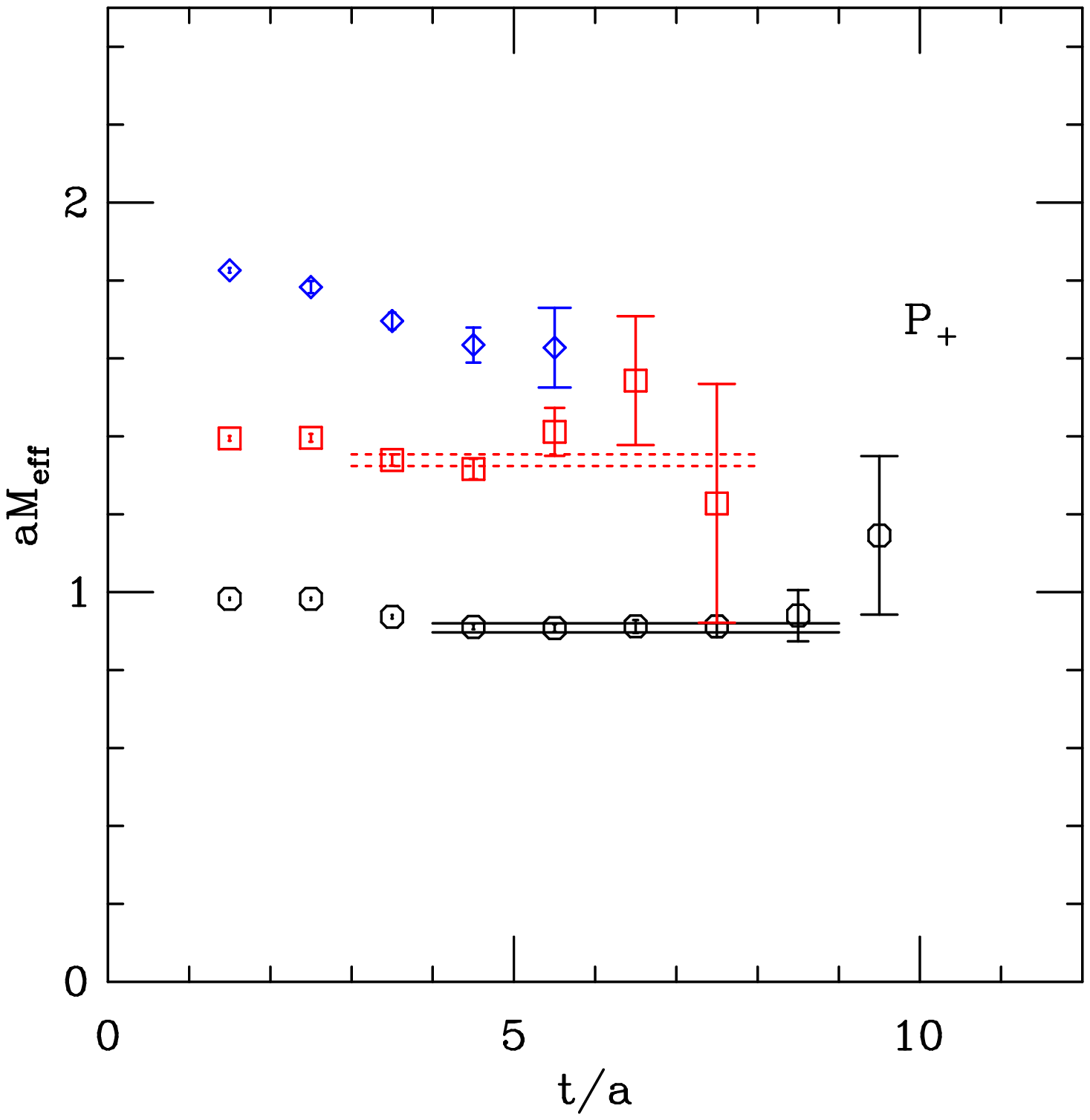}
\includegraphics*[width=3.7cm]{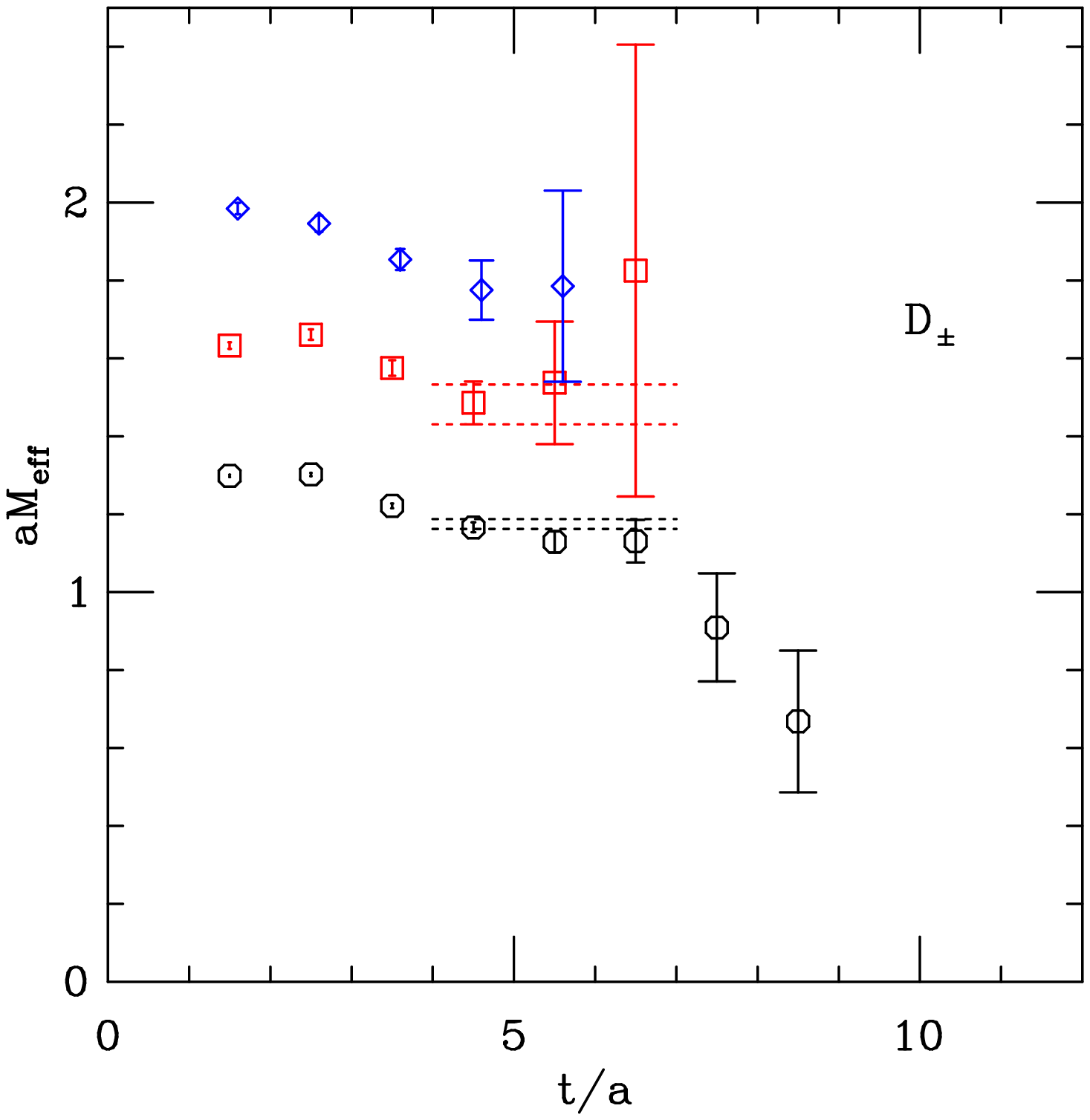}
\end{center}
\caption{
Effective masses for the static-light mesons on the quenched configurations. 
$am_q=0.08$, $a^{-1} \approx 1330$ MeV, $L \approx 1.8$ fm. 
The horizontal lines represent $M\pm\sigma_M$ fit values for the 
corresponding time ranges. 
Dashed lines indicate fits for which we adjust the minimum time for 
systematic error estimates.
}
\label{effmass}
\end{figure}

After extracting the eigenvalues, we check for single mass states by 
creating effective masses. 
A representative sample of these, along with single-elimination jackknife 
errors, are plotted against time in Fig.\ \ref{effmass}. 
In each case one finds values from the first three eigenvalues. 
The horizontal lines signify the $M\pm\sigma_M$ values which result from 
correlated fits over the corresponding range in time. 
We require that at least three effective mass points display a plateau 
(within errors) and that the eigenvectors remain constant (again, within 
errors) over the same range before we perform said fits.

\begin{figure}
\begin{center}
\includegraphics*[width=6.4cm]{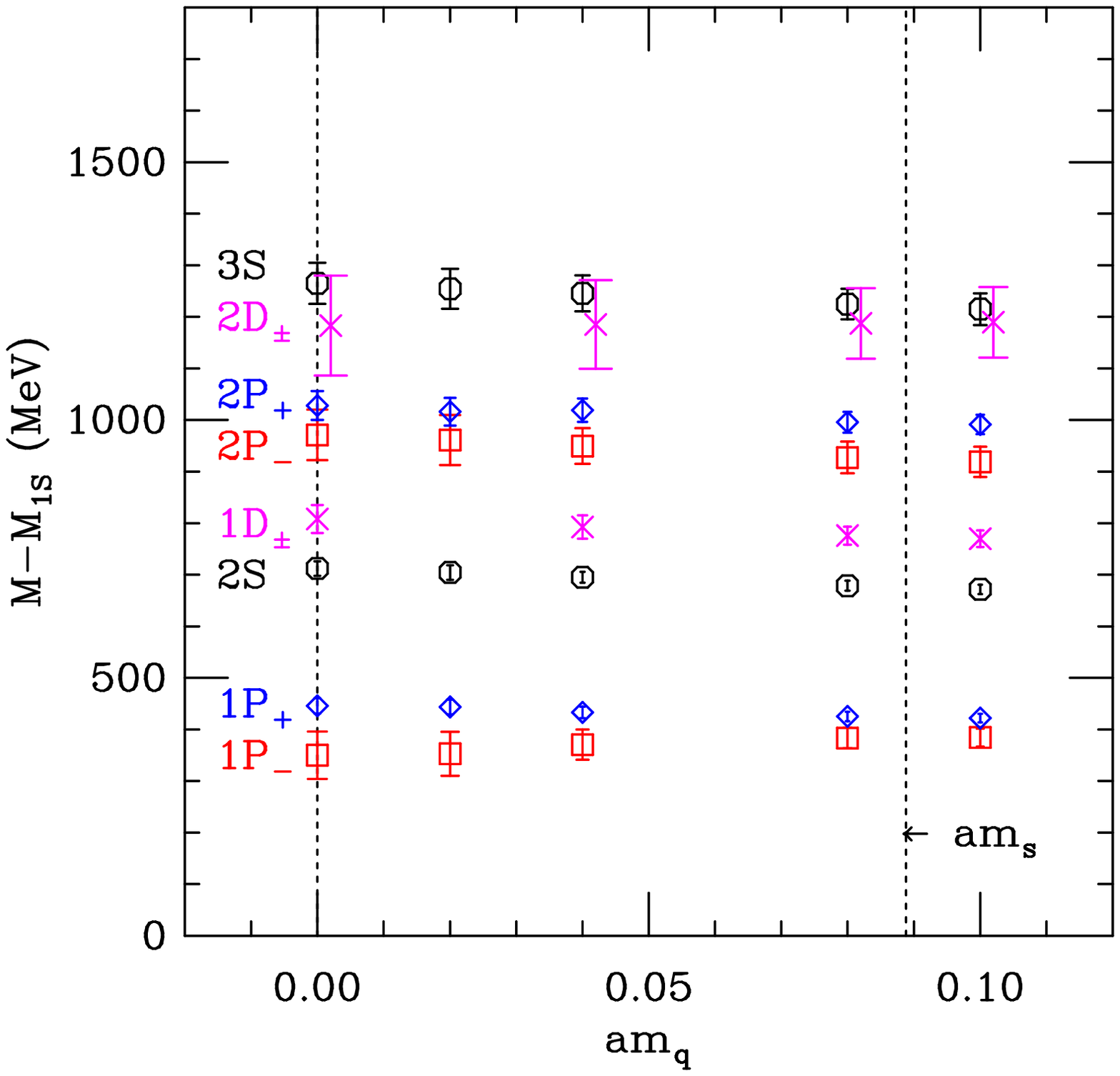}
\includegraphics*[width=6.4cm]{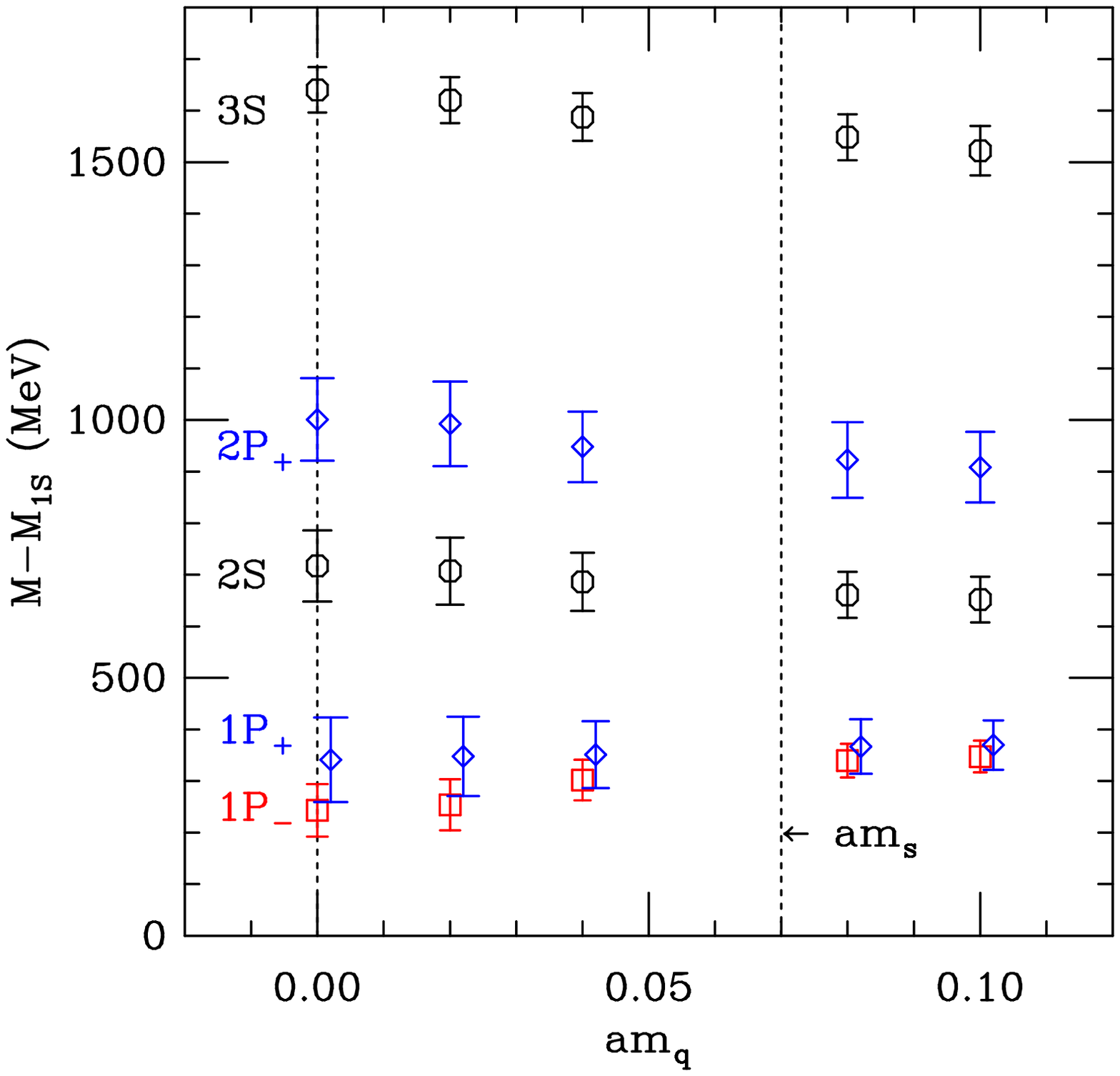}
\end{center}
\caption{
Physical mass splittings ($M-M_{1S}$) as a function of the quark mass for 
the quenched (left) and dynamical (right) lattices. 
The vertical lines denote the chiral limit ($m_q\to0$) and the strange quark 
mass ($m_s$). 
The left-most values are the (linear) chiral extrapolations.
}
\label{chiral_extrap}
\end{figure}

Performing fits for all quark masses, we next take a look at the mass 
splittings ($M-M_{1S}$) as a function of the quark mass. 
These are plotted in Fig.\ \ref{chiral_extrap}, along with the chirally 
extrapolated results ($m_q\to0$). We use simple linear fits to perform these 
extrapolations.

In Table \ref{B_split} we present the results for the 
chirally extrapolated ($B$ mesons) and strange-quark-mass interpolated 
($B_s$ mesons) mass splittings. 
We include the statistical errors from the fits in the first set of 
parentheses. 
For fits where the effective-mass plateaus are not immediately clear 
(e.g., fits represented with dashed lines in Fig.\ \ref{effmass}), 
we move the minimum time of the fit range out by 
one to two time slices and observe the subsequent changes in 
$M\pm\sigma_M$, as compared to the previous values. 
The differences from the old values are reported as systematic errors; 
these appear in the second set of parentheses. 
For a discussion of these results, we refer the reader to our lengthier 
report \cite{Burch:2006mb}.

\begin{table}
\begin{center}
\begin{tabular}{lccc} \hline
state & $J^P$ & \multicolumn{2}{c}{$M-M_{1S}$ (MeV) for $B$ / $B_s$} \\
 & & $N_f=0$, $L \approx 1.8$ fm & $N_f=2$, $L \approx 1.4$ fm \\ \hline
$2S$ & $(0,1)^-$ & 712(14) / 675(10) & 717(69) / 665(45) \\
$3S$ & $(0,1)^-$ & 1265(40)($^{+0}_{-130}$) / 1220(30)($^{+20}_{-50}$) & 1640(44)($^{+55}_{-200}$) / 1560(45)($^{+35}_{-190}$) \\ \hline
$1P_-$ & $(0,1)^+$ & 350(46) / 384(20) & 243(51) / 330(34) \\
$2P_-$ & $(0,1)^+$ & 971(49)($^{+50}_{-90}$) / 923(30)($^{+10}_{-60}$) & - / - \\ \hline
$1P_+$ & $(1,2)^+$ & 446(15) / 424(10) & 341(82) / 363(55) \\
$2P_+$ & $(1,2)^+$ & 1028(28)($^{+160}_{-80}$) / 993(20)($^{+130}_{-50}$) & 1001(80)($^{+130}_{-20}$) / 930(75)($^{+0}_{-80}$) \\ \hline
$1D_\pm$ & $(1,2,3)^-$ & 808(27)($^{+0}_{-90}$) / 773(17)($^{+0}_{-80}$) & - / - \\
$2D_\pm$ & $(1,2,3)^-$ & 1183(97)($^{+130}_{-150}$) / 1188(68)($^{+170}_{-80}$) & - / - \\ \hline
\end{tabular}
\caption{
Mass splittings for our $B$ / $B_s$ mesons. 
The first number in parentheses is the statistical error. 
The second set (if present) are the additional systematic errors which result 
from adjustments to the minimum time of the fit.
\label{B_split}
}
\end{center}
\end{table}

One thing is clear though: due to the improvement of the light-quark 
propagator estimation, and our subsequent ability to use half the points of 
the lattice as source locations, we have greatly improved our chances of 
isolating excited heavy-light states. 
In an earlier study \cite{Burch:2004aa} of heavy-light mesons using wall 
sources on the same quenched configurations, we were barely able to see the 
$2S$ state, let alone the excited states in any other operator channel. 
Also, there we used NRQCD for the heavy quark; this should only boost the 
signals since the heavy quark can then ``explore'' more of the lattice 
through its kinetic term. 
It is obvious, however, that we have much better signals now since we are 
able to see excited states in every channel ($2S$, $3S$, $2P_-$, $2P_+$, and 
$2D_\pm$) on the quenched lattice.

\acknowledgments

We would like to thank Christof Gattringer for many helpful discussions. 
We are also indebted to our colleagues in Graz -- Christian B.\ Lang, 
Wolfgang Ortner, and Pushan Majumdar -- for sharing their dynamical CI 
configurations with us. 
We also wish to thank Andreas Sch\"afer, without whose 
effort this interesting pursuit of ours would not be a paid one. 
Simulations were performed on the Hitachi SR8000 at the LRZ in Munich. 
This work is supported by GSI.

\end{document}